# Electronic Properties of Pyrochlore-Type $Ca_2Ir_2O_7$

Yuki Nakayama[1], Yoshihiko Okamoto[1,2,*], Daigorou Hirai[1], and Koshi Takenaka[1]

[1]*Department of Applied Physics, Nagoya University, Nagoya 464-8603, Japan*
[2]*Institute for Solid State Physics, University of Tokyo, Kashiwa 277-8581, Japan*

The electrical resistivity, magnetic susceptibility, and heat capacity of sintered samples of $Ca_2Ir_2O_7$, in which pentavalent Ir atoms with $5d^4$ electron configuration form a pyrochlore structure, have been studied. The obtained experimental data strongly suggest that $Ca_2Ir_2O_7$ is metallic below room temperature and exhibits no electronic or magnetic phase transitions above 0.12 K.

Pyrochlore-type iridium oxides with cubic $Fd\bar{3}m$ symmetry exhibit various remarkable electronic properties, such as a metal-insulator transition accompanied by all-in-all-out magnetic order in $Nd_2Ir_2O_7$, a possible Weyl semimetallic state in $Y_2Ir_2O_7$, and an anomalous Hall effect induced by spin chirality order in $Pr_2Ir_2O_7$.[1-4] These unique properties arise from cooperation between the strong spin-orbit coupling of $5d$ electrons, moderately strong electron correlation, and the geometry of the pyrochlore structure. Compared with these $Ir^{4+}$ oxides with $5d^5$ electron configuration, there have been few studies on the physical properties of pyrochlore-type $Ir^{5+}$ oxides with $5d^4$ electron configuration.[5-7] However, because competition between strong spin-orbit coupling and energy level splitting due to crystal field effects has been discussed in relation to $Cd_2Ir_2O_7$,[7] $Ir^{5+}$ oxides can also be expected to show interesting electronic properties. Therefore, synthesizing more pyrochlore-type iridium oxides, including pentavalent ones, and investigating their physical properties should lead to the discovery of novel electronic properties.

In this short note, we will focus on $Ca_2Ir_2O_7$ with pentavalent Ir atoms. $Ca_2Ir_2O_7$ has been synthesized by both high-pressure and hydrothermal methods, and it has been reported to crystallize in an α-pyrochlore-type form.[8,9] The magnetic susceptibility of a $Ca_2Ir_2O_7$ powder sample synthesized by the latter method showed a difference between zero-field-cooled and field-cooled data below 70 K, which, to the best of our knowledge, is the only physical property reported in previous studies.[9] Here we report the electrical resistivity, magnetic susceptibility, and heat capacity of $Ca_2Ir_2O_7$ sintered samples. These experimental data suggested that $Ca_2Ir_2O_7$ is metallic down to 0.12 K and exhibits no electronic or magnetic phase transitions.

Powder samples of $Ca_2Ir_2O_7$ were prepared by a hydrothermal method. $IrCl_3 \cdot 3H_2O$ and CaO powders in a molar ratio of 1:5 were encapsulated in a gold tube with 79 mg of $Ca(ClO_4)_2 \cdot nH_2O$ powder and 0.5 mL of pure water, and the system was kept at 1023 K and 200 MPa for 72 h.[9] The obtained samples invariably consisted of a mixture of black

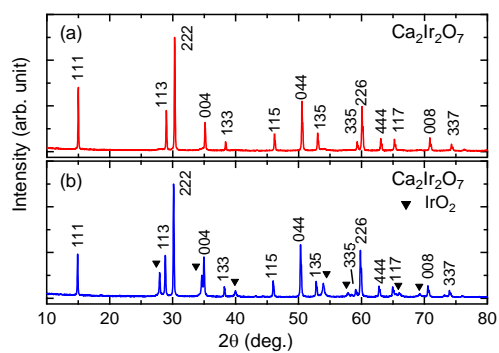

Fig. 1. Powder XRD patterns of $Ca_2Ir_2O_7$ polycrystalline samples measured at room temperature. The data taken before (a) and after (b) the sintering process are shown. The peaks indicated by the filled triangles are diffraction peaks of the $IrO_2$ impurity phase.

$Ca_2Ir_2O_7$ and white Ca hydroxide, the latter of which was removed by rinsing the sample with dilute hydrochloric acid. The obtained $Ca_2Ir_2O_7$ was pulverized, pressed into a pellet, and then sintered at 1073 K for 5 h in air. The sample was characterized by powder X-ray diffraction (XRD) analysis on a MiniFlex diffractometer (RIGAKU) employing Cu Kα radiation. Figures 1(a) and (b) show the XRD patterns of polycrystalline samples of $Ca_2Ir_2O_7$ before and after the sintering process measured at room temperature. All diffraction peaks except for small peaks of the $IrO_2$ impurity phase were indexed on the basis of a cubic unit cell with the $Fd\bar{3}m$ symmetry and lattice constants $a$ = 10.2078 and 10.2570 Å for Fig. 1(a) and (b), respectively, indicating that α-pyrochlore-type $Ca_2Ir_2O_7$ was obtained as a main phase. Since the amount of $IrO_2$ impurity appeared after the sintering process was small, it did not affect the physical property measurements. The electrical resistivity and heat capacity of the $Ca_2Ir_2O_7$ sintered sample were measured by a four-probe method and a relaxation method, respectively, using a Physical Property Measurement System (Quantum Design). The magnetization was measured using a Magnetic Property Measurement System (Quantum Design).

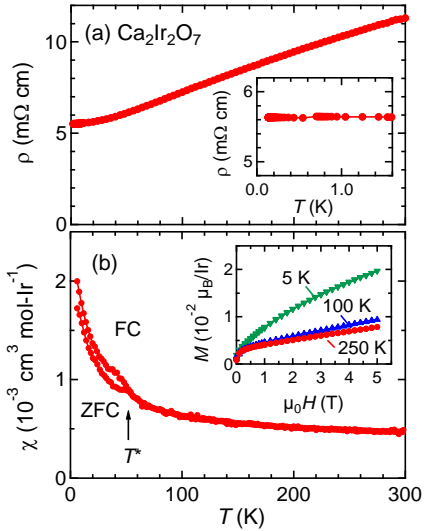

Fig. 2. Temperature dependences of (a) electrical resistivity and (b) zero-field-cooled and field-cooled magnetic susceptibility of a $Ca_2Ir_2O_7$ sintered samples. The inset in (a) shows electrical resistivity at low temperatures. The magnetic susceptibility was estimated by $\chi = [M_{5T} - M_{1T}]/\Delta H$, where $M_{5T}$, $M_{1T}$, and $\Delta H$ are magnetizations at 5 T and 1 T and a magnetic field of 4 T, respectively. The inset in (b) shows magnetization curves measured at 5, 100, and 250 K. In (b), the data after subtraction of the diamagnetic contribution from core electrons of $\chi_{dia} = -7.0 \times 10^{-5}$ cm$^3$ mol-Ir$^{-1}$ are shown.[11]

Figures 2(a, b) and 3 show the temperature dependences of the electrical resistivity, $\rho$, the magnetic susceptibility, $\chi$, and the heat capacity divided by temperature, $C/T$, for the $Ca_2Ir_2O_7$ sintered sample. The electrical resistivity was seen to decrease with decreasing temperature, suggesting metallic behavior. The residual resistivity was relatively large, as indicated by a residual resistance ratio of about 2, suggesting that the grain boundaries had a large effect on the transport properties due to the low degree of sintering. As shown in the inset of Fig. 2(a), the sample did not show a superconducting transition above 0.12 K. The heat capacity divided by temperature showed a positive value at the lowest temperature, as shown in Fig. 3. As shown in the inset, the $C/T$ data between 2 and 5 K were fitted to the equation $C/T = \beta T^2 + \gamma$, yielding Sommerfeld coefficient $\gamma$ of 19.6(7) mJ K$^{-2}$ mol-Ir$^{-1}$ and the $T^3$ term of lattice heat capacity $\beta$ of 0.89(4) mJ K$^{-4}$ mol-Ir$^{-1}$. The obtained $\gamma$ is comparable to those of metallic Ir-pyrochlore oxide $Y_{1.4}Ca_{0.6}Ir_2O_7$ and $Pb_2Ir_2O_7$, where A site is occupied by nonmagnetic ions,[6,10] although there might be a minor contribution of spin entropy of magnetic impurity, discussed later. This suggests that the ground state of $Ca_2Ir_2O_7$ is metallic, consistent with the observed $d\rho/dT > 0$ temperature dependence. However, $\gamma$ of $Ca_2Ir_2O_7$ is much larger than that of $Cd_2Ir_2O_7$,[7] which will be discussed later. The $\beta$ value yields Debye temperature of $\theta_D = 230$ K, which is lower than those for $Y_{1.4}Ca_{0.6}Ir_2O_7$ ($\theta_D \sim 400$ K) and $Pb_2Ir_2O_7$ ($\theta_D = 340$ K).[6,10] This result and the weak concave-

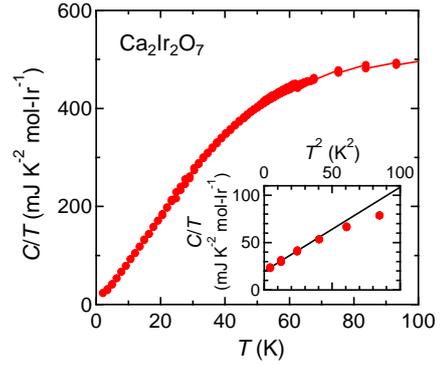

Fig. 3. Temperature dependence of heat capacity divided by temperature, $C/T$, for the $Ca_2Ir_2O_7$ sintered sample. The inset shows the $C/T$ versus $T^2$ plot, in which the solid line shows the result of linear fitting of the 2–5 K data.

downward behavior in $C/T$ versus $T^2$ plot in the inset of Fig. 3 imply the presence of a small contribution of spin entropy due to the magnetic impurity in the sample.

As shown in Fig. 2(b), there was a small difference between the zero-field-cooled (ZFC) and field-cooled (FC) magnetic susceptibilities below $T^* = 52$ K. This difference occurred at a lower temperature and was much smaller than that observed in a previous study.[9] As discussed below, it was most likely due to an extrinsic factor. The value of $\chi = 4.7 \times 10^{-4}$ cm$^3$ mol-Ir$^{-1}$ at 300 K was comparable to the Pauli paramagnetic susceptibility, $\chi_{Pauli}$, estimated by using $\gamma = 19.6$ mJ K$^{-2}$ mol-Ir$^{-1}$ and the Wilson ratio, $R_W$. The strongly correlated limit of $R_W = 2$ gives $\chi_{Pauli} = 5.3 \times 10^{-4}$ cm$^3$ mol-Ir$^{-1}$ and the weakly correlated limit of $R_W = 1$ gives $\chi_{Pauli} = 2.7 \times 10^{-4}$ cm$^3$ mol-Ir$^{-1}$. Although it is difficult to estimate the strength of electron correlation from these $\chi$ and $\chi_{Pauli}$ values, this result suggests that the Pauli paramagnetism has a considerable contribution in $\chi$ data at high temperatures. The increase of $\chi$ at low temperatures might not be due to the change of $\chi_{Pauli}$, because the magnetization curve at 5 K is concave downward between 1 and 5 T in contrast to the linear behavior at 100 and 250 K.

The absence of an anomaly at $T^* = 52$ K in the $\rho$ and $C/T$ data strongly suggested that the difference between ZFC and FC magnetic susceptibilities below $T^*$ had an extrinsic origin. If this difference was associated with long-range magnetic ordering of the Ir spins, such as in the 5d pyrochlore oxides $Cd_2Os_2O_7$ and $Nd_2Ir_2O_7$,[1,12,13] an anomaly would appear at $T^*$ in the electrical resistivity and heat capacity data. However, as can be seen in Figs. 2(a) and 3, there was no anomaly at $T^* = 52$ K in the $\rho$ or $C/T$ data, indicating that the difference between ZFC and FC $\chi$ must be extrinsic, potentially due to magnetic impurities. In other words, these results suggested that $Ca_2Ir_2O_7$ does not show magnetic order and remains in a paramagnetic state.

Finally, we will discuss the electronic state and spin-orbit coupling in $Ca_2Ir_2O_7$. In Ir oxides with IrO$_6$ octahedra, the effect of strong spin-orbit coupling is often prominent in determining the electronic properties. In the strong spin-orbit coupling limit, an $Ir^{5+}$ oxide becomes a nonmagnetic insulator with a gap between electronic states with effective total angular momenta of $J_{eff}$ = 3/2 and 1/2.[14] In fact, $Cd_2Ir_2O_7$ showed a semimetallic electronic state with a small density of states at $E_F$, as indicated by γ = 0.6 mJ $K^{-2}$ $mol$-$Ir^{-1}$. It is argued that this semimetallic state was realized by mixing of $J_{eff}$ = 3/2 and 1/2 bands near $E_F$ due to the trigonal distortion of an $IrO_6$ octahedron and the wide Ir $5d$ band.[7] In contrast, $Ca_2Ir_2O_7$ showed metallic experimental data with a large γ of 19.6 mJ $K^{-2}$ $mol$-$Ir^{-1}$, suggesting the possibility that the trigonal distortion of an $IrO_6$ octahedron in $Ca_2Ir_2O_7$ was larger than that in $Cd_2Ir_2O_7$. Although further structural parameters, including the site occupancies related to chemical deficiencies, should be determined in future studies, these results clearly indicated that the pentavalent Ir oxide with pyrochlore structure is promising for controlling electronic properties over a wide range from metallic to insulating.

## ACKNOWLEDGMENTS

The authors are grateful to T. Yamauchi for his support with the magnetic susceptibility measurements. The work was partly carried out under the Visiting Researcher Program of the Institute for Solid State Physics, the University of Tokyo and partly supported by JSPS KAKENHI (Grant Numbers: 19H05823, 20H02603).

*e-mail: yokamoto@issp.u-tokyo.ac.jp